\documentclass[a4paper,10pt,oneside]{article}

\usepackage{amsmath,amssymb,amsfonts}%
\usepackage{amsthm}%
\usepackage{booktabs}%

\usepackage[utf8]{inputenc}
\usepackage[T1]{fontenc}
\usepackage[english]{babel}

\usepackage{bbold}
\usepackage{braket} 

\title{The quantum-classical bipartite universe and the logical emergence of $\hbar$ from classical temporality}
\author{Vincenzo Chilla}
\date{}

\begin{document}

\maketitle

\begin{abstract}

We present a foundational reinterpretation of quantum dynamics based on the standard von Neumann formalism, positing a structural logical bipartition between a classical observational environment and a quantum observed system. Against universalist frameworks, we argue that theoretical consistency requires distinct propositional logics: Boolean logic governs the environment, while non-Boolean logic applies to the observed system. Within this framework, an ideal classical clock is represented as an open system whose energy spectrum is uniformly discretized. The dynamics follow from the Gorini-Kossakowski-Sudarshan-Lindblad (GKSL) master equation under the constraint that the Hamiltonian remains observable. Although the clock's temporal evolution is driven by dissipative coupling to the measurement apparatus, it admits a deterministic solution corresponding to information-preserving cyclic dynamics. This deterministic regime may emerge as a logical attractor, stabilized by the Boolean structure inherent to the measurement interaction. Ultimately, $\hbar$ is derived as a fundamental scaling parameter from the classical clock's action operator, revealing that the quantum of action reflects the discrete logical structure underlying classical temporality. Furthermore, this framework establishes an exact mapping with 't Hooft's cellular automaton formalism, offering an alternative ontological interpretation that not require hidden variables and superdeterministic assumptions. The resulting distinction between quantum dynamics (as potential recontextualization) and classical dynamics (as temporal actualization) provides a novel logical approach to the measurement problem. 



\end{abstract}

\section{Introduction}
\label{intro}

Physics is not just the study of nature; rather, it is the study of nature \emph{as it can be known and logically communicated}~\cite{Heisenberg1958,Bohr1961}. It is meaningless to ask what the ``quantum world'' is like independently of observation, precisely because the requirement of observability \emph{and} logical communicability is a fundamental prerequisite for what constitutes a physical theory. This is the epistemological reason why a \emph{fundamental} physical theory must necessarily and explicitly incorporate the logical structure of the observational environment, in relation to the observed object, as one of its constitutive elements. Thus, the ``weirdness'' of quantum mechanics should not be considered a flaw in the theory; rather, it is the hallmark of a physical reality observed in its potentiality, awaiting actualization through the ``transcendental'' condition for logical communication among observers: the act of measurement.

In light of the premise set forth above, the formalization of quantum mechanics developed by J.~von Neumann in his celebrated treatise~\cite{vNeumann1955}, which remains the standard formalism taught today, is well suited to serve as the essential core of a fundamental physical theory. This formalization does not merely provide a physical description of the quantum world; it also presupposes the logical structure required to distinguish, within the observational universe -- which is epistemologically \emph{closed} by definition -- between the physics of the observational environment and that of the observed object~\cite{BirkhoffvNeumann1936,Landsman2017}. The former is governed by a Boolean experimental logic. The latter, by contrast, is governed by a non-Boolean experimental logic until it is incorporated into the environment itself through the logico-physical act of measurement. In other words, the former is strictly \emph{classical}, while the latter is strictly \emph{quantum}.

Despite this duality inherent in the formalism, von Neumann's framework has typically been given a universalist interpretation, beginning with the author himself~\cite{Laudisa2024}. That is, one assumes either that physical reality is fundamentally described by quantum logic, with classicality emerging as a phenomenological regime in which the quantum of action is negligible relative to the action of the observed system, or, less commonly, that physical reality is fundamentally described by classical logic, with quantumness emerging as an epistemic regime in which the quantum of action is \emph{not} negligible relative to the action of the observed system. Both perspectives, however, present a conceptual problem with respect to the closure of the observational universe: they cannot properly include the observer in the theory without generating inconsistencies~\cite{Zinkernagel2015,FrauchigerRenner2018}. In fact, in the former case, it is impossible to express the logical meaning of measurement (the ``problem of measurement''~\cite{Margenau1963}); in the latter case, the physical requirement of measurement independence -- i.e., the possibility of independently designating the observables to be measured -- is compromised, leading to the ``problem of determinism'' or, at least, ``superdeterminism''~\cite{Bell2004}.

In this work, within the observational universe as described by the von Neumann framework, we explicitly adopt the \emph{logico-physical bipartition} between the observer and the observed system -- the so-called \emph{Heisenberg cut}~\cite{Heisenberg1930,BacciagaluppiCrull2009} -- and examine its role in the temporal dynamics of the two regimes. Indeed, the \emph{movability} of this cut, as highlighted by decoherence experiments~\cite{MKTKIWSMW2000,HUHRBZA2003}, corresponds to a shift in the dynamical regime: from the determinism of unitary evolution, governed by the Schr\"odinger equation, to the indeterminism of measurement completion dictated by the reduction postulate. This very movability serves as the bridge between the logical establishment of the \emph{measurement context} (through the expression of measurement independence) and the actual physical emergence of the resulting classical (i.e., Boolean) regime as an act of environmental incorporation and ontic determination. That is, once the system has been appropriately prepared within a given context and the corresponding cut has been fixed, the observational universe will evolve deterministically and unitarily until a new measurement is independently completed and the cut is repositioned.\footnote{Here, \emph{independently} means that the repositioning of the cut is logically, rather than causally, connected to the evolution of the observational universe. It is important to emphasize that logic does not necessarily imply ``consciousness''.}

The two partitions of the observational universe correspond to two distinct conceptions of \emph{time}, which nevertheless converge at the cut. On the one hand, there is a \emph{quantum} conception, in which the passage of time marks a potential transition of the observational context -- namely, a \emph{recontextualization} of the universe~\cite{BCGKL2022}, understood as the exploration of alternative Boolean environments through unitary evolution. On the other hand, there is a \emph{classical} conception, in which the same temporal flow, as registered by a \emph{clock} within a \emph{fixed} observational context, reveals the unfolding -- i.e., the dynamics -- of its actual physical reality up to the completion of the measurement at each \emph{tick}. A prototypical case of this dynamical duality has recently been identified and thoroughly investigated by G.~'t Hooft in his cellular automaton model for the harmonic oscillator~\cite{tHooft2024}. Although 't Hooft's original interpretation relies on ``hidden variables'', we intend to show in this paper that the same dynamical duality finds its natural interpretation within the bipartite framework adopted here, without the need for additional ontological structures. It is important to emphasize that this dynamical evolution is fully deterministic and involves no change in the information content of the observational universe. It is only upon the completion of a measurement that information undergoes a change, thereby establishing the ``arrow of time'' and the ``physical present''~\cite{Thebault2022}.

The paper is structured as follows. In Section~\ref{vnframe}, we review the main features of von Neumann's Hilbert-space framework, emphasizing the distinction between the classical observational environment and the quantum observed object. Particular attention is devoted to the role of the classical clock and to the ``problem of stationarity'' in closed classical systems. In Section~\ref{dyn}, we contrast quantum and classical dynamics with respect to their conceptual treatment of ontic states and epistemic accessibility. This discussion focuses on closed systems, employing the Koopman-von Neumann formalism for classical dynamics~\cite{KoopmanvNeumann1932,Bondar2012}. In Section~\ref{opendyn}, we consider classical dynamics formulated in Hilbert space for open systems. This analysis proves crucial for providing classical systems with a time evolution consistent with the general formalism of quantum mechanics. In Section~\ref{clock}, we concentrate on open classical systems with equally spaced Hamiltonian spectra, which are particularly significant in modeling ideal classical clocks as an essential part of the observational environment; their Fourier analysis is also presented to elucidate this structure. Based on this analysis and the assumption of a fundamental logical bipartition of the observational universe, the quantum of action is defined in terms of the structural parameters of an ideal classical clock. In Section~\ref{hidden}, we recast 't Hooft's cellular automaton as an ideal classical clock rather than as a hidden-variable model, thereby avoiding the need for superdeterminism while resolving difficulties in extending the model to non-equidistant spectra. Finally, in Section~\ref{concl}, we present our concluding remarks and discuss potential implications.

\section{Von Neumann's bipartite framework}
\label{vnframe}

In this section, we review the standard non-relativistic Hilbert-space formalism of von Neumann~\cite{vNeumann1955} in its salient features concerning the quantum-to-classical transition, referring the reader to~\cite{Chilla2026} for further details. 

\subsection{Logical bipartition and the observational universe}
\label{logbip}

It should first be emphasized that, as shown in~\cite{Chilla2026}, the von Neumann formalism possesses a foundational universal character in the sense that its mathematical structure is applicable to both the quantum and classical realms. However, this universality does not imply a monistic physical interpretation. Rather, imposing the observer-observed duality -- namely, the partition between classical and quantum realities -- within the formalism entails a logical bipartition (with consequent physical implications). The propositional logic pertaining to the observational environment must be Boolean to guarantee the communicability of the observed reality among observers (this is the \emph{epistemic} sense of objectivity and classicality à la Bohr). Conversely, the propositional logic pertaining to the observed object is generally described by a non-Boolean orthomodular lattice, which captures the meaning of ontological potentiality prior to the logical disambiguation conferred by measurement completion~\cite{BirkhoffvNeumann1936}. The consequent physical implication concerns the \emph{preparable states} represented by means of density matrices: the aforementioned requirement of Booleanity corresponds to the mutual \emph{commutativity} of the preparable states of the observational environment. While the von Neumann formalism typically requires the commutativity of observables, as shown in~\cite{Chilla2026}, the fundamental commutativity concerns the states; consequently, commutativity between observables follows from spectral theory. Such commutativity is, on the other hand, not required for the preparable states of the observed object. This logical approach to classicality clearly differs from the standard one, in which classical physics emerges from quantum physics as a dynamical limit ($\hbar \to 0$). In fact, in this approach, a system can be classical even when its action is on the order of $\hbar$, regardless of whether they are subsequently amplified and recorded.

The fundamental physical structure described by the von Neumann formalism is thus the observational universe, which is closed by definition and, by the evolution postulate, always subject to unitary dynamics unless a measurement occurs. When one studies the time evolution of a quantum system $S$, it is implicitly assumed that there exists a \emph{classical clock} that registers the events of measurement completion, formalized by the reduction postulate~\cite{BGL1994,AHB2022}. Therefore, this clock is not an external, abstract concept or parameter but a \emph{physical system that belongs to a pre-existing observational environment} $A$.\footnote{This is a fundamental tenet of the theory of relativity. Even a system that implements a ``quantum clock'' presupposes a classical clock for the logical objectivity of the timekeeping.} Specifically, if $t=0$ represents the instant of the last measurement completion and $t=T$ the instant of the subsequent one -- as registered by the clock in the environment -- the observational universe evolves for $ t \in (0,T)$ according to the von Neumann equation (evolution postulate)
\begin{equation}
i \hbar \frac{d\rho_{AS}}{dt} =  [\hat{H}_{AS}, \rho_{AS}],
\end{equation}
where $\rho_{AS}$ represents the state of the bipartite universe consisting of the observational environment $A$ and the observed object $S$. From the postulate on composite systems, the universal Hamiltonian, denoted by $\hat{H}_{AS}$, can be written in the form 
\begin{equation}
 \hat{H}_{AS} = \sum_{j} \ket{\epsilon_j} \bra{\epsilon_j} \otimes \hat{V}^{(j)},
 \label{hasgen}
\end{equation}
which realizes the logico-physical bipartition discussed above, ensuring the commutativity of the environmental states by construction, as the environmental projectors are invariant under this Hamiltonian. Here, the vectors $\ket{\epsilon_j}$ (the ``pointer states'') form the basis of the environment's Hilbert space $\mathcal{H}_A$ -- which we term the \emph{classical basis} -- consisting of the eigenvectors common to the preparable states. The operators $\hat{V}^{(j)}$ represent appropriate observational potentials acting on the Hilbert space $\mathcal{H}_S$ of $S$ where the total Hilbert space of the universe is given by $\mathcal{H} = \mathcal{H}_A \otimes \mathcal{H}_S$.\footnote{Here, we assume, for simplicity, that the dimension of the state space of the observable universe is finite. The decomposition $\mathcal{H} = \mathcal{H}_A \otimes \mathcal{H}_S$ formalizes the conceptual notion of the Heisenberg cut.} These potentials correspond to the conditional unitary evolution of $S$: if the environment is assumed to be in the state labeled by $j$, the potential $\hat{V}^{(j)}$ acts as the evolution Hamiltonian of the observed system. In fact, it readily follows that, once the environmental state is assumed to be $\ket{\epsilon_j} \bra{\epsilon_j}$, we can write 
\begin{equation}
\rho_{AS} = \ket{\epsilon_j} \bra{\epsilon_j} \otimes \rho_S^{(j)}, \label{mist}
\end{equation} 
and the conditional state of $S$, $\rho_S^{(j)}$, acting on $\mathcal{H}_S$, also satisfies the von Neumann equation in the form
\begin{equation}
i \hbar \frac{d\rho_{S}^{(j)}}{dt} =  [\hat{V}^{(j)}, \rho_{S}^{(j)}].
\label{vneumann}
\end{equation} 

Consequently, what the physicist typically calls ``the closed quantum system $S$ with Hamiltonian $H$'' is in fact only a subsystem of the complete bipartite structure, with state space $ \mathcal{H}=\mathcal{H}_A \otimes \mathcal{H}_S $, under a specific potential $\hat{V}^{(j)} \equiv \hat{H}$. The presumed unitary evolution of $S$ is well-defined only because the environment $A$ provides the classical background (the clock) that paces this evolution. In other words, \emph{considering the unitary dynamics of a given quantum system $S$ also presupposes a classical environment $A$ (in an actualized physical state) that completes the observational universe and provides the fundamental clock}. For this reason, the standard universalist approach fails when extending the quantum description to the entire universe. It neglects the classical observational environment prescribed by the formalism, specifically the universal Hamiltonian in the block-diagonal form given in~\eqref{hasgen}, which is mandated by the logico-physical bipartition and by unitary temporal evolution paced by a physical clock. This block-diagonal structure is a standard mathematical tool in decoherence theory~\cite{Zurek2003,Zurek2009}, where it appears as part of the interaction Hamiltonian that selects the preferred pointer basis. However, within our bipartite framework, its physical interpretation diverges from the standard description. While standard decoherence relies on system-environment entanglement to explain the emergence of classicality from a purely quantum substrate, our approach posits that this block-diagonal structure fundamentally describes the physical \emph{relocation} of the Heisenberg cut. Crucially, \emph{this very transition presupposes a pre-existing, logically Boolean environment corresponding to the original position of the cut, which provides the classical background necessary to govern and register the dynamics}.

\subsection{Physical states and observability}
\label{stobs}

Building on the logico-physical bipartition established above, we note that since the vectors of the classical basis constitute the \emph{ontic} states of the classical environment $A$ (as required by its Boolean logic), the vectors of any other basis in $\mathcal{H}_A$ have only formal mathematical significance and cannot be physically prepared or measured. Therefore, quantum superpositions of classical basis states are excluded from the set of physically admissible states for $A$. This exclusion is not a dynamical consequence of environmental tracing, but rather a fundamental constraint imposed by the logical requirement of Booleanity. Such a kinematic restriction confines the physical state space of the environment to statistical mixtures of classical basis states
\begin{equation}
\rho_A = \sum_j p_j \ket{\epsilon_j}\bra{\epsilon_j},
\label{rida}
\end{equation}
where the Kolmogorovian probabilities $p_j$ (with $p_j \geq 0$ and $ \sum_j p_j =1 $) carry an exclusively \emph{epistemic} meaning, representing the uncertainty about which ontic state the environment actually occupies. Generalizing Eq.~\eqref{mist} to the case where the environment is in such a statistical mixture, the form of the physical state of the observational universe becomes
\begin{equation}
\rho_{AS} = \sum_j p_j \ket{\epsilon_j} \bra{\epsilon_j} \otimes \rho_S^{(j)}. 
\label{mist1}
\end{equation} 
Recall that, in sharp contrast to the environment $A$, for a given ontic state $\ket{\epsilon_j}\bra{\epsilon_j}$ of the environment, the conditional state of system $S$, $\rho_S^{(j)}$, exhibits a genuine \emph{ontic indeterminacy} during its unitary evolution, which only measurement completion can resolve. As a consequence, the quantum probabilities characterizing such a state are not of an epistemic Kolmogorovian nature, and the corresponding density matrix is generally non-diagonal, reflecting the quantum superpositions that characterize the state of $S$.

Von Neumann's formalism postulates that every admissible observable is represented by a Hermitian operator; however, not all Hermitian operators are admissible observables.\footnote{Note that physical states are also represented by Hermitian operators -- specifically, density operators -- yet they are not technically considered observables.} In the case of a classical system, the eigenstate-eigenvalue link requires that an observable yield definite values in all physically preparable states. This dictates that \emph{any admissible observable $\hat{O}$ must commute with all physically preparable states $\rho$ of the system} (\emph{observability condition}):
\begin{equation}
[ \hat{O}, \rho ] = 0.
\label{obsc}
\end{equation} 
The observability condition implies that all admissible observables mutually commute and are simultaneously diagonal in the classical basis, thereby eliminating the contextuality inherent to the quantum framework.\footnote{In his impossibility proof for hidden variables~\cite{vNeumann1955}, von Neumann presupposed that any linear combination of Hermitian operators constitutes an observable with correspondingly linear expectation values -- a premise J.~Bell later deemed unwarranted for hidden-variable frameworks~\cite{Bell1966}. In fact, since such theories operate within single-contextual (classical) domains rather than quantum multi-contextuality, linearity is only required to hold among commuting observables compatible with joint observability conditions.} This enables us to distinguish between two different characterizations of observables, a \emph{strong} and a \emph{weak} one, depending on whether the framework is exclusively classical or bipartite. Given that in the purely classical framework the context is fixed, a Hermitian operator representing an admissible observable must be strictly diagonal in the classical basis. In contrast, in the bipartite framework, where recontextualization in time is permitted, a Hermitian operator represents an admissible observable only if \emph{there exists a context} -- that is, a specific dynamical positioning of the Heisenberg cut -- in which it commutes with the physically preparable states.\footnote{Throughout this paper, we adopt the following notational convention: operators representing admissible observables (in either the strong or weak sense) are denoted with a hat (e.g., $\hat{O}$, $\hat{x}$, $\hat{p}$, $\hat{H}$), whereas operators that do not represent observables -- such as the Liouvillian operator or auxiliary generators -- are written without a hat (e.g., $L$, $\lambda_x$, $\lambda_p$).}
 
Now, as applied to the quantum framework, the evolution postulate explicitly requires that the Hamiltonian of a system be an admissible observable (generating the unitary evolution of the closed system). Since the \emph{logical} constraint of classicality -- i.e., the mutual commutativity of preparable states -- pertains to the states rather than the fundamental definition of an observable itself, the Hamiltonian remains an admissible observable in the classical framework as well, and thus in the strong sense defined above. In other words, the environmental context remains \emph{essential}; here, ``essential'' denotes the dynamical context defined by the observables that commute with the Hamiltonian. However, the reduced dynamics of the observational environment, as obtained from the von Neumann equation for the closed observational universe, results in trivial evolution for $\rho_A$ in light of the observability condition~\eqref{obsc} and the specific form of the Hamiltonian~\eqref{hasgen}, namely \emph{stationarity}. This result is hardly surprising, as \emph{von Neumann evolution for a closed system in the quantum framework simply entails potential recontextualization}; since recontextualization is precluded in the classical framework, \emph{the dynamics of a classical closed system necessarily becomes trivial}. 

The consequence for a classical clock embedded in the environment is immediate: given that such a clock must register the events of measurement completion, it cannot be a closed system if it is to pace the recontextualization dynamics of the quantum system. Indeed, while the classical background of the environment must remain stationary to guarantee objectivity, the clock itself must function as an open subsystem, exchanging information with the classical measuring apparatus to undergo the distinguishable state changes required to mark the passage of time. We will analyze the clock dynamics in detail in Sections~\ref{opendyn} and \ref{clock}; however, before doing so, a brief review of closed quantum dynamics and standard classical dynamics in phase space within the Koopman-von Neumann formalism is in order (see Section~\ref{dyn}).

\section{Closed quantum and classical dynamics}
\label{dyn}

While the unitary evolution of closed quantum systems is universally governed by the von Neumann equation, the classical counterpart is traditionally described in phase space via the Liouville equation. The Koopman-von Neumann (KvN) formalism bridges this conceptual divide by recasting classical Hamiltonian mechanics into a Hilbert-space framework, wherein the Liouville equation assumes a von Neumann-like form~\cite{KoopmanvNeumann1932,Bondar2012}. Within the bipartite framework outlined above, it is both useful and instructive to examine the quantum dynamics of the closed system $S$, while explicitly highlighting that the Koopman-von Neumann formalism is generally unsuited for describing the environment $A$. Indeed, the KvN framework inherently presupposes classicality as a strict dynamical condition -- effectively the $\hbar \to 0$ limit, with its attendant phase-space structure -- whereas in the bipartite framework, classicality is a logical prerequisite (Booleanity) that is prior to and independent of any dynamical limit or phase-space description. Although the KvN formalism is traditionally interpreted as realizing the $\hbar \to 0$ limit, we argue that, since the Boolean structure of the environment's propositional logic cannot guarantee the dynamical condition $\hbar \to 0$ assumed by KvN, the latter formalism fails to capture the logical nature of classicality as defined in our framework. Furthermore, it is here highly advantageous to formulate the dynamics of $S$ within the Heisenberg picture rather than the standard Schr\"{o}dinger picture, thereby aligning with the emphasis in classical mechanics on the temporal evolution of observables.

Consider a closed quantum system $S$ described by the Hilbert space of states $\mathcal{H}_S = L^2(\mathbb{R})$ and the standard Hamiltonian corresponding to one-dimensional motion along $x$:
\begin{equation}
\hat{H} = \frac{\hat{p}^2}{2m} + V(\hat{x}),
\label{hams}
\end{equation}
where the potential $V$ is represented by an analytic function on $\mathbb{R}$. Working within the Heisenberg picture, the equations of motion for the observables $\hat{x}$ and $\hat{p}$ are given by:
\begin{equation}
i \hbar \frac{d\hat{x}}{dt} = -[\hat{H}, \hat{x}],
\label{heisx}
\end{equation}
\begin{equation}
i \hbar \frac{d\hat{p}}{dt} = -[\hat{H}, \hat{p}].
\label{heisp}
\end{equation}
Substituting the Hamiltonian from~\eqref{hams} into~\eqref{heisx} and~\eqref{heisp}, and recalling the commutation relation $[\hat{x}, \hat{p}] = i \hbar$, one immediately obtains the equation:
\begin{equation}
m \frac{d^2 \hat{x}}{dt^2} = -V'(\hat{x}),
\label{newton}
\end{equation}
which formally corresponds to the classical Newtonian equation of motion. This correspondence between the evolution equation for the quantum observable $\hat{x}$ and the classical Newtonian equation is not accidental and confirms that the algebraic structure of the equations of motion is independent of the dynamical limit -- indeed, $\hbar$ disappears from the operator equation -- highlighting that classicality is rooted in the algebraic, rather than dynamical, structure of the theory. However, this correspondence does not imply that quantum effects are nonexistent; rather, they manifest themselves in the evolution equation for the expectation values of $\hat{x}$, which are directly tied to the measured dynamical quantities. Specifically, due to the non-commutativity of quantum operators, the equation for $\langle \hat{x} \rangle$ does not reduce to the simple classical trajectory equation $m\frac{d^2\langle \hat{x}\rangle}{dt^2} = -V'(\langle \hat{x}\rangle)$, because $\langle V'(\hat{x})\rangle \neq V'(\langle \hat{x}\rangle)$ due to higher-order correlation terms arising from the Taylor expansion of $V'(\hat{x})$. It is precisely at the level of expectation values, and not at the operator level, that the quantum nature of the dynamics becomes apparent, in agreement with the logical framework outlined in Section~\ref{vnframe}, where classicality is defined as a Boolean property of the observational environment rather than as a dynamical limit. 

It is worth emphasizing that the quantum Hamiltonian~\eqref{hams}, defined on $L^2(\mathbb{R})$, can never represent a classical system due to the commutation relation $[\hat{x}, \hat{p}] = i\hbar \neq 0$. Indeed, $\hat{x}$ and $\hat{p}$ cannot simultaneously be observables of a single (essential) context, since strong observability requires mutual commutativity with all physically preparable states. Consequently, the very form of the Hamiltonian loses its classical meaning. Within the KvN framework, to obtain a classical description corresponding to the Hamiltonian~\eqref{hams}, one must impose the commutativity $[\hat{x}, \hat{p}] = 0$ and change the state space to $L^2(\mathbb{R}^2)$, i.e., \emph{phase space}, where $\hat{x}$ and $\hat{p}$ act as multiplication operators
\begin{equation}
    \hat{x} \, \psi(x,p) \doteq x \psi(x,p), \quad \hat{p} \, \psi(x,p) \doteq p \psi(x,p), \quad \text{with } \psi \in L^2(\mathbb{R}^2).
\end{equation}
Significantly, \emph{this transition does not represent the presumed dynamical limit ($\hbar \to 0$) but rather a logical restructuring}: the imposition of commutativity is precisely the requirement of Booleanity for the algebra of observables, which defines classicality as a logical property rather than a dynamical regime. By introducing the auxiliary (unobservable) generators $\lambda_x$ and $\lambda_p$ on $L^2(\mathbb{R}^2)$, defined as
\begin{equation}
    \lambda_x \doteq -i \frac{\partial}{\partial x}, \quad \lambda_p \doteq -i \frac{\partial}{\partial p},
\end{equation}
such that
\begin{equation}
    [ \hat{x}, \lambda_x] = [ \hat{p}, \lambda_p] = i,
\end{equation}
one can write the Liouvillian operator 
\begin{equation}
    L \doteq \frac{\partial \hat{H}}{\partial p} \lambda_x - \frac{\partial \hat{H}}{\partial x} \lambda_p = i \{ \hat{H}, \cdot \},
\end{equation}
where $\{ \cdot , \cdot \} $ denotes the ordinary Poisson bracket. In this way, $L$ serves as the generator of the dynamics for any classical observable $\hat{O}(\hat{x},\hat{p})$ as an operator on $L^2(\mathbb{R}^2)$:\footnote{In the Schrödinger picture, the evolution equation for the classical state $\rho(x,p)$ is the Liouville equation $\frac{d \rho}{d t} = -i [L, \rho ] = \{ \hat{H}, \rho \}$. This equation is the classical counterpart of the von Neumann equation, in which the Liouvillian operator $L$ plays the role of the observable $\hat{H} / \hbar$ as the generator of temporal evolution. }
\begin{equation}
\frac{d\hat{O}}{dt} = i [ L,\hat{O}] = \{ \hat{O}, \hat{H} \}.
\end{equation} 
In particular, from the dynamical equations for $\hat{x}$ and $\hat{p}$, and using the Hamiltonian in the form of~\eqref{hams}, one precisely recovers the Newtonian equation of motion as in~\eqref{newton}, but now in its classical form, with $\hat{x}$ acting on the phase space $L^2(\mathbb{R}^2)$.

However, the logical restructuring highlighted above entails a significant transformation in the very nature of the dynamics. In the quantum framework, dynamics is generated by the Hamiltonian $\hat{H}$, which is an \emph{observable} in the sense defined in Section~\ref{vnframe} -- that is, it commutes with the physically preparable states and is therefore admissible within a given observational context. The unitary evolution governed by $\hat{H}$ thus corresponds to a \emph{potential recontextualization} of the observational universe in the sense that the system evolves by exploring the space of alternative observational contexts, each associated with a different Boolean environment, until a measurement actualizes one of these contexts. This is the essence of quantum dynamics -- an evolution in potentiality, awaiting actualization through measurement. In sharp contrast, once the Booleanity constraint is imposed and the classical phase-space description is adopted, the dynamics is no longer generated by the Hamiltonian $\hat{H}$ (which, as an observable, would induce recontextualization), but rather by a Liouvillian operator $L = i \{\hat{H}, \cdot\}$, which is \emph{not} an observable in the sense of Section~\ref{vnframe}. Indeed, $L$ does not commute with the physically preparable states of the classical system, and therefore cannot be admitted as an observable within any fixed Boolean context. This is not a mere technicality: it reflects the fact that classical dynamics is \emph{not} a recontextualization -- since the classical context is fixed and essential by the very definition of Booleanity -- but rather a \emph{temporal unveiling} of the system's actual existence. The Liouvillian governs the epistemic disclosure of what is already ontically determined within the fixed classical context, rather than the exploration of alternative contexts.

This fundamental distinction explains why the dynamics of a closed classical system is necessarily \emph{ontically trivial} (or \emph{logically stationary}): since the classical framework is strictly fixed within a single context -- that is, it admits no recontextualization by definition -- and since recontextualization is precisely what an observable Hamiltonian generates, the only consistent dynamics within a fixed Boolean context is one generated by a non-observable Liouvillian. But such dynamics, being merely an unveiling of actual existence rather than a recontextualization, cannot produce any genuine change in the ontic state of the system. Hence, the stationarity of closed classical dynamics is not an accident or a limitation of the formalism, but rather a logical necessity arising from the Boolean nature of the classical observational environment. The logical priority of Booleanity over dynamics is thus manifest: it is the logical constraint of classicality that determines the form of the dynamics, not vice versa. This fundamental distinction will prove crucial in the following section, where we show how classical systems are treated as open systems in order to exhibit non-trivial dynamics consistent with ordinary phase-space evolution.

\section{Open classical dynamics} 
\label{opendyn} 

To derive non-trivial classical dynamics, we start from the general framework for open systems, e.g., quantum systems described by arbitrary finite-dimensional Hilbert spaces interacting with an environment. The most general equation governing the evolution of the quantum state -- namely, the evolution of the associated density matrix, which must satisfy the physical requirements of \emph{complete positivity} and \emph{trace preservation} -- is the Gorini-Kossakowski-Sudarshan-Lindblad (GKSL) equation~\cite{GKS1976,Lindblad1976}. This provides the general form of a \emph{time-local} (Markovian) master equation capable of generating a completely positive and trace-preserving (CPTP) dynamical map for all times, and is therefore of fundamental significance in the treatment of open systems. Without detailing its derivation, for which we refer the reader to the standard literature~\cite{BreuerPetruccione2007,ChruscinskiPascazio2017}, the GKSL equation generalizes the von Neumann equation as follows:
\begin{equation} 
    i \hbar \frac{d \rho}{d t} = [\hat{H}, \rho] + i \hbar \sum_k \left( L_k \rho L_k^\dagger - \frac{1}{2} L_k^\dagger L_k \rho - \frac{1}{2} \rho L_k^\dagger L_k \right), 
    \label{gksl} 
\end{equation} 
where the Lindblad operators $L_k$ represent the jump operators associated with the information flow induced by the system-environment coupling. 

As established in Section~\ref{dyn}, a closed classical system exhibits trivial (stationary) dynamics because its Hamiltonian must commute with its physical states. However, by treating the classical system as an open system and imposing this same classicality constraint within the GKSL equation~\eqref{gksl}, the unitary term vanishes ($[\hat{H}, \rho] = 0$), and we obtain non-trivial dynamics describing the interaction of the classical system with its environment:
\begin{equation} 
  \frac{d \rho}{d t} = \sum_k \left( L_k \rho L_k^\dagger - \frac{1}{2} L_k^\dagger L_k \rho - \frac{1}{2} \rho L_k^\dagger L_k \right) \equiv \mathfrak{L}[\rho].  
  \label{gksl-class} 
\end{equation}
Note that the explicit dependence on $\hbar$ disappears from the dynamical equation~\eqref{gksl-class} not as a result of the $\hbar \to 0$ limit, but rather due to the strong observability condition for $\hat{H}$. Consequently, the dynamical equation~\eqref{gksl-class} acquires a \emph{purely classical} character, and the Lindblad operators $L_k$ -- although dependent on the specific physical modeling -- \emph{need not contain $\hbar$}. This allows $\hbar$ to be introduced subsequently as a scale parameter of the classical model under consideration. However, self-consistency requires that this ``$\hbar_{\textrm{classical}}$'' coincides with the universal constant $\hbar$ appearing in the quantum evolution postulate. By contrast, in the classical regime understood as the quantum dynamical limit ($\hbar \to 0$), $\hbar$ represents a fundamental quantity and the Lindblad operators generally retain an implicit $\hbar$-dependence.

The classicality constraint requires that the state $\rho$ be diagonal in the classical basis $\{ \ket{\gamma_k} \}$, composed of the eigenstates of $\hat{H}$ describing the system; however, this is not required for the Lindblad operators $L_k$, which are not observables. Instead, they must be elementary jump operators between states of the classical basis, labeled by indices $i,j$ ($i \neq j$):
\begin{equation} 
    L_{ij} = \sqrt{W_{ij}} \ket{\gamma_i} \bra{\gamma_j},
    \label{lindw} 
\end{equation}
where $W_{ij} \geq 0$ is the transition rate from state $j$ to state $i$. The operators $L_{ij}$, for all possible pairs $(i,j)$ with $i \neq j$, form a complete set of Lindblad operators that preserve the diagonality required by classicality. Therefore, the Lindbladian $\mathfrak{L}[\rho]$ in~\eqref{gksl-class} can be rewritten as
\begin{equation} 
    \mathfrak{L}[\rho] = \sum_{i\neq j} \left( L_{ij} \rho L_{ij}^\dagger - \frac{1}{2} L_{ij}^\dagger L_{ij} \rho - \frac{1}{2} \rho L_{ij}^\dagger L_{ij} \right).   
\end{equation}
Recalling that the state $\rho$ can be expressed as a statistical mixture in the form $\rho = \sum_k p_k \ket{\gamma_k}\bra{\gamma_k}$ and using Eq.~\eqref{lindw}, we can explicitly compute the various contributions to $\mathfrak{L}[\rho]$:\footnote{These contributions are associated with the so-called \emph{gain} and \emph{loss} terms relative to the population of the classical states.}
\begin{align}
    L_{ij} \rho L_{ij}^\dagger &= W_{ij} p_j \ket{\gamma_i}\bra{\gamma_i}, \\
    \frac{1}{2} L_{ij}^\dagger L_{ij} \rho + \frac{1}{2} \rho L_{ij}^\dagger L_{ij} &= W_{ij} p_j \ket{\gamma_j}\bra{\gamma_j}, \\
    L_{ij}^\dagger L_{ij} &= W_{ij} \ket{\gamma_j}\bra{\gamma_j}.
\end{align}
Therefore, summing over $i$ and $j$ with $i \neq j$, the Lindbladian becomes
\begin{align}
    \mathfrak{L}[\rho] &= \sum_{i \neq j} \left( W_{ij} p_j \ket{\gamma_i}\bra{\gamma_i} - W_{ij} p_j \ket{\gamma_j}\bra{\gamma_j} \right) \nonumber \\
    &= \sum_i \left( \sum_{j \neq i} W_{ij} p_j - \sum_{j \neq i} W_{ji} p_i \right) \ket{\gamma_i}\bra{\gamma_i}.
    \label{pau}
\end{align}
Since the time derivative of the state is given by
\begin{equation}
    \mathfrak{L}[\rho] = \frac{d \rho}{d t} = \sum_i \frac{d p_i}{d t} \ket{\gamma_i}\bra{\gamma_i},
\end{equation}
comparison with the previous expression in~\eqref{pau} yields the well-known Pauli master equation~\cite{Davies1974}:
\begin{equation}
    \frac{d p_i}{d t} = \sum_{j \neq i} \left( W_{ij} p_j - W_{ji} p_i \right).
    \label{pauli}
\end{equation}

The Pauli master equation in~\eqref{pauli} can also be written in the matrix form
\begin{equation}
    \frac{d \vec{p}}{d t} = \mathfrak{W} \vec{p},
    \label{pauli1}
\end{equation}
where $\vec{p}$ represents the column vector of probabilities $p_i$, and $\mathfrak{W}$ is the transition rate matrix whose elements $\mathfrak{W}_{ij}$ are given by $W_{ij}$ for $i \neq j$ (so the off-diagonal elements are non-negative) and $\mathfrak{W}_{ii} = -\sum_{j \neq i} W_{ji}$, ensuring probability conservation (i.e., the elements of each column sum to zero). The formal solution of Eq.~\eqref{pauli1} is given by
\begin{equation}
    \vec{p}(t) = e^{\mathfrak{W} t} \vec{p}(0) \equiv \mathfrak{T}(t) \vec{p}(0),
    \label{paulisol}
\end{equation}
where the matrix $\mathfrak{T}(t) = e^{\mathfrak{W} t}$ is a \emph{stochastic} matrix representing a continuous-time Markov chain. It is worth noting that, although the Hamiltonian $\hat{H}$ does not drive the dynamics (given the observability condition, there is no recontextualization), its eigenvalues $E_i$ remain physically relevant and are implicitly encoded in the stochastic evolution of $\mathfrak{T}(t)$ through the transition rates $W_{ij}$. In particular, if the classical system under consideration is coupled to a thermal bath at temperature $T$, the transitions between states must satisfy the \emph{detailed balance condition} to ensure thermal equilibrium:
\begin{equation}
    \frac{W_{ij}}{W_{ji}} = \frac{p_i^{\mathrm{eq}}}{p_j^{\mathrm{eq}}} = e^{-\frac{E_i - E_j}{k_B T}}.
    \label{det_balance}
\end{equation}

From a foundational perspective, it is essential to emphasize that the randomness inherent in the transition rates $W_{ij}$ is purely \emph{epistemic}. At a fundamental ontological level, the total composite system -- comprising the classical system and its classical environment -- evolves in a strictly deterministic unveiling. The apparent stochasticity captured by the transition matrix $\mathfrak{T}(t)$ does not imply fundamental indeterminism, as recontextualization is precluded. Rather, such stochasticity emerges as a direct consequence of tracing out the environmental degrees of freedom: this partial trace operation acts as a coarse-graining procedure over the underlying deterministic dynamics, yielding an effective, reduced description of the system in which the inaccessible microscopic information of the environment manifests as epistemic probabilistic transitions.

\section{The ideal classical clock}
\label{clock}

Having established the general framework for classical dynamics in Hilbert space as a logical constraint -- rather than a dynamical limit -- derived from open quantum systems, this section addresses the specific problem of temporality within this formalism. Building on the standard quantum clock constructions of Salecker and Wigner~\cite{SaleckerWigner1958} and Peres~\cite{Peres1975}, we formulate the \emph{ideal clock} within our classical operatorial framework. Thus, an ideal classical clock is defined as a \emph{classical open system} characterized by a Hamiltonian $\hat{H}_C$ with an \emph{equally spaced} energy spectrum, coupled to a measurement apparatus acting as its environment. The equally spaced structure -- corresponding to a translational symmetry of the Hamiltonian eigenstates -- is preserved as an effective property inherited from the quantum model, ensuring temporal regularity despite the absence of coherent quantum evolution. This classical formulation follows naturally from the framework: since the clock is necessarily embedded in the classical description obtained via the strong observability constraint and partial trace, it is conceptually consistent to describe it within the classical, stochastic framework. We note in passing that the mathematical structure developed below parallels 't Hooft's cellular automaton model~\cite{tHooft2024}, though its physical interpretation is different, as discussed in Section~\ref{hidden}. Consequently, the derivation presented here leverages this existing formal structure to provide a descriptive, conceptual exposition tailored to our bipartite framework. 

\subsection{General stochastic dynamics}
\label{clockdyn}

As shown in Section~\ref{opendyn}, the ideal classical clock evolves according to the Pauli master equation~\eqref{pauli}, where the transition rates $W_{nm}$ encode the spectral properties of $\hat{H}_C$. Assuming that the transition rate $W_{nm}$ depends solely on the energy difference $E_n - E_m$ between levels $n$ and $m$, the equally spaced nature of the spectrum,
\begin{equation}
E_n = E_0 + n\Delta E ,
\end{equation} 
implies that the rate becomes a function of the index difference alone; we denote this function by $W(j)$, so that $W_{nm} = W(n-m)$. Introducing the descriptive time step $\delta t $ -- understood here as an epistemic parametrization rather than a fundamental temporal interval -- and restricting the clock's state space to a finite dimension $ N \gg 1 $, the column-stochastic matrix $\mathfrak{T}$ defined in~\eqref{paulisol} can be expanded to first order in $\delta t$ as follows:
\begin{equation}
    \mathfrak{T}_{nm}(\delta t) = 
    \begin{cases}
        W(n-m)\,\delta t & \text{for } n \neq m, \\
        1 - \displaystyle\sum_{k \neq n} W(k-n)\,\delta t & \text{for } n = m .
    \end{cases}
    \label{tcirc}
\end{equation}
The expression~\eqref{tcirc} is a first-order discretization: it represents a stochastic matrix when $\delta t$ is sufficiently small that its entries remain nonnegative. Since the rates depend only on the index difference, $\mathfrak{T}(\delta t)$ is \emph{circulant}. This circulant structure arises from the \emph{periodic boundary conditions} imposed on the system, $W(j) \equiv W(j \bmod N)$, which can be interpreted as an information-preserving reset, thereby closing the $N$-level clock cycle without loss of temporal information content. Therefore, here and below, indices are understood modulo $N$. 

The matrix $\mathfrak{T}(\delta t)$ acts on the probability vector $\vec{p} \in \mathbb{R}^N$. It is not, by this fact alone, a physical unitary operator on $\mathcal{H}_C$. Nevertheless, as an $N\times N$ matrix, it can also be represented as a linear operator on the $N$-dimensional vector space spanned by $\{\ket n\}$. In this representation, its circulant structure allows it to be diagonalized by the discrete Fourier transform:
\begin{equation}
    \ket{\varphi_k}
    \doteq
    \frac{1}{\sqrt{N}}\sum_{n=0}^{N-1}
    e^{-in\varphi_k}\ket n,
    \qquad
    \varphi_k=\frac{2\pi k}{N},
    \label{fourier}
\end{equation}
where $k\in\{0,\dots,N-1\}$. We refer to $\{\ket{\varphi_k}\}$ as the \emph{Fourier basis}. Here it is a \emph{formal} basis entailing no empirically observable consequences. Indeed, an operator diagonal in the Fourier basis does not commute with $\hat{H}_C$, and therefore does not constitute an admissible observable in the sense of Section~\ref{vnframe}. In any case, the introduction of the Fourier basis proves useful for describing the evolution of the clock as a dynamic epistemic unveiling of its ontic state, as we shall see in the following. The eigenvalues of the circulant matrix $\mathfrak{T}(\delta t)$ in~\eqref{tcirc} are given by 
\begin{equation} 
    \mu_k = \sum_{m=0}^{N-1} \mathfrak{T}_{m0} (\delta t) e^{i m \varphi_k}  . 
\end{equation} 
Substituting the matrix elements from Eq.~\eqref{tcirc}, we obtain: 
\begin{equation} 
    \mu_k = 1 + \delta t \sum_{m=0}^{N-1} W(m) (e^{i m \varphi_k} -1) \doteq 1 + \lambda_k \delta t, 
    \label{eigen}
\end{equation} 
where $\lambda_k$ is the characteristic rate associated with the mode $\ket{\varphi_k}$. In particular, $\lambda_0=0$, as required by conservation of total probability. 

The continuous-time Markov generator associated with the rates is
\begin{equation}
    \mathfrak{W}_{nm}
    =
    \begin{cases}
        W(n-m), & n\ne m, \\
        -\displaystyle\sum_{k\ne m}W(k-m), & n=m.
    \end{cases}
\end{equation}
Thus,
\begin{equation}
\mathfrak{T}(\delta t)=I+\delta t\,\mathfrak{W}+O(\delta t^2).
\end{equation}
In the limit of small time steps at fixed rates, the corresponding continuous-time evolution is generated by $e^{t\mathfrak{W}}$. The eigenvalues of $\mathfrak{W}$ are $\lambda_k$, and its spectral representation, as a matrix acting on the probability space, is
\begin{equation}
    \mathfrak{W}
    =
    \sum_{k=0}^{N-1}\lambda_k
    \ket{\varphi_k}\bra{\varphi_k}.
    \label{liuvstoc}
\end{equation}
This is a matrix representation of the Markov generator; it does not identify $\mathfrak{W}$ with an observable on $\mathcal{H}_C$. Note that, for a Markov generator, the eigenvalues satisfy $\operatorname{Re}\lambda_k\leq 0$. We may write
\begin{equation}
\lambda_k=-\alpha_k+i\omega_k,
\qquad
\alpha_k\geq 0,
\end{equation}
reflecting the generally dissipative and irreversible nature of the classical clock's dynamics. Specifically, the coefficients $\alpha_k$ represent the relaxation rates of the system, while the coefficients $\omega_k$ represent the oscillation frequencies of the Fourier modes. 

\subsection{Deterministic discrete-time regime}
\label{nonstoc}

We now introduce an ideal \emph{deterministic} discrete-time clock model in which the one-step evolution is defined directly as the \emph{cyclic permutation}
\begin{equation}
    T(\delta t)
    \doteq
    \sum_{n=0}^{N-1}
    \ket{(n+1)\bmod N}\bra n.
    \label{shift_op}
\end{equation}
As a transition matrix acting on probability vectors, $T(\delta t)$ is deterministic and stochastic: each state is mapped to exactly one successor state. As a linear operator on the auxiliary space $\mathbb{C}^N$, it is unitary and acts on the classical basis as
\begin{equation}
    T (\delta t)\ket n=\ket{(n+1)\bmod N}.
    \label{shift}
\end{equation}
With the Fourier convention in Eq.~\eqref{fourier}, its eigenvalues are
\begin{equation}
    T (\delta t) \ket{\varphi_k}=e^{i\varphi_k}\ket{\varphi_k},
    \qquad
    \varphi_k=\frac{2\pi k}{N} .
    \label{fourier_shift}
\end{equation}
Thus, the cyclic permutation in the classical basis is represented by phase multiplication in the Fourier basis. The phases are the $N$th roots of unity, as required for a one-step cyclic shift.

This model parallels the general stochastic model described in Section~\ref{clockdyn} in the particular case in which the relaxation rates vanish, $\alpha_k = 0$, and the Fourier mode frequencies are integer multiples of the fundamental frequency:
\begin{equation}
\omega_k \doteq \frac{\varphi_k}{\delta t} = \omega_1 k.
\end{equation}
Then, one may define the cycle duration
\begin{equation}
\tau \doteq N\delta t, 
\end{equation}
which represents the time scale of the deterministic dynamics, i.e., the total time required for the system to explore all $N$ energy levels before closing the cycle with a physical \emph{tick}. The fundamental frequency is thus 
\begin{equation}  
    \omega_1 = \frac{\varphi_1}{\delta t} = \frac{2 \pi}{N \delta t}  = \frac{2 \pi}{\tau} ,
\end{equation} 
and the interval $\tau$ defines the intra-cycle time within which the clock retains information prior to the next ontic readout. In this way, \emph{the provided dual description -- cyclic permutation in the classical basis $\{ \ket{n} \}$ and phase rotation in the Fourier basis $\{ \ket{\varphi_k} \}$ -- exposes a discretized phase space where these bases serve as conjugate variables}. The clock dynamics updates one energy level per step $\delta t$ with perfect Fourier coherence, embodying the regular, noise-free progression of classical temporality as the zero-noise limit of the stochastic framework.

We observe that, in general, the physical stability of the deterministic regime ($\alpha_k=0$) presents a fundamental challenge in Markovian frameworks, where cyclic permutations are typically unstable under generic dynamical perturbations~\cite{SaloffZuniga2010}. In our formulation, however, we propose that stability can be obtained not from dynamical fine-tuning but from the logical bipartition between quantum system and classical environment. Thus, the deterministic regime described above is stabilized by the Boolean environment, which serves as a structural constraint rather than a dissipative thermal bath: periodic boundary conditions implement an \emph{information-preserving reset} that constrains the transition matrix $\mathfrak{W}$ to the anti-Hermitian subspace.\footnote{In other words, the cyclic solution persists under perturbations that preserve the Boolean structure and information distinguishability of minimal projectors. The induced evolution remains an automorphism (a permutation in finite models). This does not imply asymptotic attraction or robustness to stochastic mixing; it expresses the \emph{persistence of the deterministic cyclic sector} under transformations conserving the defining logical conditions of the classical clock.} In fact, the deterministic clock regime constitutes a logically protected sector -- i.e., a \emph{logical} attractor -- in which $\alpha_k = 0$ follows from the non-dissipative nature of the measurement interaction itself, ensuring robust harmonic isochronism by structural necessity rather than parameter tuning.    

The deterministic shift can be represented on diagonal density matrices by conjugation with $T(\delta t)$. If
$ \rho(t)=\sum_{j=0}^{N-1}p_j(t)\ket j\bra j$,
then
\begin{equation}
\begin{aligned}
    \rho(t+\delta t) &= T(\delta t)\,\rho(t)\,T(\delta t)^\dagger \\
    &= \sum_{j=0}^{N-1}p_j(t)\,
       \ket{(j+1)\bmod N}\bra{(j+1)\bmod N},
\end{aligned}
\end{equation}
which can be written in terms of populations,
\begin{equation}
    p_{j+1}(t+\delta t)=p_j(t),
    \label{probevol}
\end{equation}
with indices modulo $N$. This confirms that probabilities translate at each interval $\delta t$ as the epistemic unveiling of the clock's state is realized. Similarly, a clock observable $\hat O(t)=\sum_{j=0}^{N-1}o_j(t)\ket j\bra j$ is represented in the Heisenberg picture by
\begin{equation}
\begin{aligned}
    \hat O(t+\delta t) &= T(\delta t)^\dagger\hat O(t)T(\delta t) \\
    &=  \sum_{j=0}^{N-1}o_{j+1}(t)\ket j\bra j,
    \end{aligned}
\end{equation}
so that
\begin{equation}
    o_j(t+\delta t)=o_{j+1}(t),
\end{equation}
again with indices modulo $N$. This describes the corresponding cyclic relabelling of the epistemic readout of the observable over time. 

\subsection{Continuum deterministic limit and emergence of $\hbar$}
\label{contlim}

As in~\cite{tHooft2024}, taking the continuum limit $N \to \infty$ while keeping the cycle time $\tau = N \delta t$ and the spectral spacing $\Delta E$ fixed reshapes the mathematical structure of the clock without altering its physical interpretation. In this limit, $\delta t\to 0$ and the finite set of clock states is replaced, at the level of the state space, by the countably infinite set $\{\ket{n}:n\in\mathbb{N}_0\}$, whose Hilbert-space completion is $\ell^2(\mathbb{N}_0)$. The Hamiltonian is represented as
\begin{equation}
    \hat{H}_C
    =
    E_0\mathbb{1}_C+\Delta E\,\hat{n},
    \qquad
    \hat{n}
    =
    \sum_{n=0}^{\infty}n\ket n\bra n ,
    \label{hclock}
\end{equation}
where the constant term $E_0\mathbb{1}_C$ contributes only an overall energy offset and will be omitted where it plays no role. As $N\to\infty$, the discrete Fourier modes become dense in phase. 

Formally, analogously to~\eqref{fourier}, one introduce the phase-labelled generalized vectors:
\begin{equation}
    \ket{\varphi}
    \doteq
    \frac{1}{\sqrt{2\pi}}
    \sum_{n=0}^{\infty}e^{-in\varphi}\ket n,
    \qquad
    \varphi\in[0,2\pi),
    \label{phase_vectors}
\end{equation}
which however are not normalizable elements of $\ell^2(\mathbb{N}_0)$; they are to be understood as a formal Fourier representation.\footnote{Although this construction is related to phase-state formalisms such as those of Susskind and Glogower~\cite{SusskindGlogower1964} and Pegg and Barnett~\cite{PeggBarnett1989}, its present use is intended as a mathematical representation of the clock, not as a claim that $\varphi$ corresponds to an admissible observable.} The classical clock states $\rho (t)$ therefore take the form: 
\begin{equation}
\rho (t) = \int_0^{2\pi} p(\varphi,t) \ket{\varphi} \bra{\varphi} d \varphi,
\end{equation} 
with $p(\varphi,t) \geq 0$ and $ \int_0^{2\pi} p(\varphi,t) d \varphi = 1$; and, as in the discrete case described in~\eqref{probevol}, the clock probability distribution $p(\varphi,t) $ evolves by rotating at the characteristic frequency $\omega_1 = 2 \pi / \tau$: 
\begin{equation}
p(\varphi, t) = p (\varphi - \omega_1 t , 0).
\label{clockevol}
\end{equation}
Significantly, despite the Hamiltonian $\hat{H}_C$ in~\eqref{hclock} mirroring the form of a harmonic oscillator, the temporal evolution of the clock is strictly governed by the continuous rotation acting on the \emph{classical} basis vectors. The point is that the base $\{ \ket{\varphi} \}$ ($ \varphi \in [0, 2 \pi) $) and the basis $\{ \ket{n} \}$ ($n \ge 0$) are connected by the \emph{Fourier duality}, while rotation reproduces the time evolution of the energy eigenstates of a harmonic oscillator: 
\begin{equation}
\ket{n} (t) = e^{-i \omega_1 n t} \ket{n} (0).
\label{qhoevol}
\end{equation}

We can define a \emph{characteristic action scale} $\hbar_c$ from the finite-clock parameters. Ignoring the irrilevant constant energy offset $E_0$, we introduce the \emph{action} operator $\hat{J}$ as
\begin{equation}
    \hat{J}
    \doteq
    \frac{\hat{H}_C-E_0\mathbb{1}_C}{\omega_1}
    =
    \frac{\Delta E}{\omega_1}\hat{n}
    =
    \hat{n}\frac{\tau\Delta E}{2\pi}.
    \label{action_scale}
\end{equation}
Then, the quantity
\begin{equation}
    \hbar_c
    \doteq
    \frac{\tau\Delta E}{2\pi}
    \label{hclass0}
\end{equation}
has dimensions of action, and Eq.~\eqref{action_scale} becomes $\hat J=\hbar_c \hat n$. We now claim that \emph{the caracteristic action $\hbar_c$ must be exactly the same constant as in the quantum evolution postulate}, i.e., the \emph{Planck's constant} $\hbar$:\footnote{This definition is consistent with the Bohr-Sommerfeld quantization condition $\oint p\,dq = nh$, yielding $J = n\hbar$ for periodic systems with equally spaced spectra~\cite{Sommerfeld1916}.}
\begin{equation}
\hbar \equiv \hbar_c.
\label{hclass1}
\end{equation} 
In fact, the identification in~\eqref{hclass1} follows as a consequence of the bipartite structure of the observational universe. To see this, consider the \emph{classical} Hamiltonian of the clock (up to the offset), $\hat{H}_C =  \omega_1 \hat{J} = \hbar_c \omega_1 \hat{n}$, promoted to an observable of the entire bipartite observational universe, where $S$ denotes the \emph{quantum} observed system:
\begin{equation}
    \hat{H}_C \otimes \mathbb{1}_S = \left( \hbar_c \omega_1 \sum_{n=0}^\infty n \ket{n}\bra{n} \right) \otimes \mathbb{1}_S.
\end{equation}
Now, recalling Eq.~\eqref{hasgen}, we reinterpret the observable $\hat{H}_C \otimes \mathbb{1}_S$ as the Hamiltonian of a \emph{dual} observational universe:
\begin{equation}
\hat{H}_C \otimes \mathbb{1}_S = \sum_{n=0}^\infty \ket{n}\bra{n} \otimes \left( \hbar_c \omega_1 n \, \mathbb{1}_S \right).
\end{equation}
In this dual picture, the \emph{classical} environment is described by the same clock (pointer) basis $\{ \ket{n} \}$, and the \emph{quantum} system $S$ is subject to an observational potential:
\begin{equation}
    \hat{V}^{(n)} = \hbar_c \omega_1 n \, \mathbb{1}_S.
\end{equation} 
This potential corresponds precisely to that of a quantum system, observed with respect to the clock, which evolves undergoing potential recontextualization with a fundamental frequency $\frac{\hbar_c}{\hbar}\omega_1$, as established by the unitary evolution postulate. On the other hand, as discussed above and in particular as shown in Eqs.~ \eqref{clockevol} and \eqref{qhoevol}, its temporal actualization -- as paced by the clock -- is characterized by a fundamental frequency $\omega_1$. Thus, to ensure a uniforme quantum-classical temporality -- i.e., the same fundamental frequency -- the characteristic action $\hbar_c$ must coincide with the quantum of action $\hbar$:
\begin{equation}
\hbar \equiv \frac{\tau\Delta E}{2\pi}.
\label{hclass}
\end{equation} 
We emphasize that this identification is a derived \emph{consistency requirement} imposed by the bipartite structure of the observational universe and the unitary evolution postulate, which requires a uniform temporality.  

As anticipated in Section~\ref{opendyn}, this definition of $\hbar$ -- implied by the pure classical clock dynamics -- is consistent with: (i) a notion of classicality as a logical prerequisite rather than a dynamical limit ($\hbar \to 0$); and (ii) the assumption of unitary quantum evolution for isolated systems, which underlies the derivation of the GKSL equation~\eqref{gksl}. In this way, the classical (logical) environment unveils the actualized quantum state of the observed object at each ``tick of the ideal classical clock'', transforming the continuous evolution into a \emph{discrete, irreversible sequence of events} occurring at regular time intervals $\tau$ -- the cyclic ticks of the clock upon measurement completion. Thus, the ideal classical clock ensures that the dynamics is not a chaotic succession of random jumps, but a structured, periodic progression that can serve as a reliable time standard.

\section{The cellular automaton as a classical clock}
\label{hidden}

In Section~\ref{clock}, we described the dynamics of the ideal classical clock. In its mathematical formulation, this system inherently involves an equally spaced energy spectrum and periodic boundary conditions, thereby fully recovering the formalism employed by 't Hooft in his ``cogwheel model''~\cite{tHooft2016}. This also aligns with the duality~\cite{tHooft2024} between a deterministic (classical) \emph{cellular automaton} -- which evolves ``at the beat of the clock''~\cite{tHooft2016} -- and the (quantum) harmonic oscillator. The linchpin of this duality is the exact mapping between the ontological basis, which describes the ontic state of the system, and its Fourier-transformed basis, which plays a formal role in the epistemic characterization of the dynamics. As 't Hooft argues, this reveals how classical hidden variables may work in general: the classical states form the ontological basis underlying the quantum description.

However, the bipartite framework for the observational universe, adopted in the present work, admits an interpretation of the formalism that does not necessitate hidden variables. Contrary to the premises of the ``cogwheel model'', the ontological basis is precisely represented by the eigenstates $\ket{n}$ of the clock Hamiltonian $\hat{H}_C$. Indeed, $\hat{H}_C$ constitutes the genuine observable (in the strong sense) of the system, which is situated as part of the classical observational environment and is therefore necessarily Boolean. The Fourier basis $\{ \ket{\varphi_k} \}$ of this true classical basis, conversely, serves merely as a formal instrument for the description of the classical dynamics, as established in Section~\ref{clock}. Consequently, whereas from the perspective of the ``cogwheel model'' the ontological basis is ``hidden'' by the Fourier basis that generates the ``strange'' quantum dynamics, within the bipartite framework, the ontological basis is ``revealed'' by the Fourier basis that yields the harmonic isochronism of the classical clock. 

The proposed interpretation recasts the cellular automaton as an ideal classical clock model rather than a hidden-variable theory, thereby overcoming two key difficulties: (i) extension to systems with non-equidistant energy spectra -- since the clock structure is tied to the Boolean environment rather than to a specific Hamiltonian of the observed system; and (ii) reliance on superdeterminism to evade Bell's theorem constraints on universal classical theories -- since the bipartition between classical environment and quantum system preserves measurement independence by construction. This reframing establishes the cellular automaton as a fundamental classical anchor of the observational framework, resolving the associated interpretational paradoxes without invoking hidden variables or superdeterminism.

\section{Discussion and outlook}
\label{concl}

A foundational approach to physics that aspires to universalism entails an inescapable epistemological limitation: quantum universalism lacks a logico-physical description of the observer system; conversely, classical universalism lacks a logico-physical description of the measured system. Only a bipartition of the universe can account for the physical coexistence of both. While a subsystem of the universe may be quantum or classical at two distinct instants -- namely, at two distinct ticks of a classical clock -- the universe as a whole cannot logically be entirely quantum at any given instant. The standard von Neumann postulates are fully consistent with this fundamental bipartition of the universe, providing the formalism to describe both the observer system, treated as classical, and the measured system, treated as quantum. Crucially, within this framework, the unitary evolution of a system inherently presupposes a classical clock within the universe to parameterize its temporal dynamics. Conversely, a discontinuous and non-unitary evolution -- as required by the projection postulate -- is equally necessary for the definition of a \emph{tick}.

The unitary dynamics of a quantum system carries a fundamentally different theoretical significance from its classical counterpart: while the former describes the physical process of potential recontextualization -- governed by the system's Hamiltonian, which is an observable -- the latter pertains to the logical process of unveiling the current contextual ontic state -- governed by the system's Liouvillian, which is \emph{not} an observable. The conceptual bridge between these two meanings resides in the definition of the quantum of action, $\hbar$, which acts as a universal constant for the dynamics of the quantum system and, consistently, as a finite and non-vanishing fundamental scaling parameter for the dynamics of an ideal classical clock. Consequently, in a dynamical regime where the theories of gravitation, special relativity, and quantum mechanics are of comparable significance and the relevant energy scale approaches the Planck energy, the quantization of action entails the discreteness of the temporal dimension (Planck time). This discreteness of time is elegantly captured by 't Hooft's cogwheel formalism, which, ticking at the Planck frequency, governs the mechanics of the observational universe at a fundamental level.

The necessity of a classical clock that objectively demarcates the dynamics of quantum phenomena is particularly evident in Einstein's conception of physics, notably during the debates over the completeness of quantum theory. In fact, the celebrated photon-box thought experiment, explicitly devised by Einstein to challenge the validity of the time-energy uncertainty relation, relies on the concept of an ideal classical clock. The tick period, $\tau$, must be less than or equal to the duration of the box's opening interval. Furthermore, the clock's energy uncertainty, quantified by $\Delta E$, inevitably stems from its unavoidable exchanges with the surrounding classical environment. For the experiment to yield a precise measurement, this uncertainty must be sufficiently small to allow the energy of the emitted photon to be known with the desired precision. It follows from~\eqref{hclass} that, contrary to the loophole Einstein sought to exploit with this specific setup, this level of precision cannot be arbitrarily increased, regardless of the method used to determine the photon's energy. While Bohr famously resolved this paradox, invoking gravitational time dilation, our approach reveals a deeper origin. This confirms that the time-energy uncertainty relation arises fundamentally from the structural and operational constraints of the ideal classical clock and, in contrast to the position-momentum case, does not carry the same ontological implications, as time remains a parameter rather than an observable.

Paradoxically, the intrinsically epistemic character of the time-energy uncertainty relation can only reinforce Einstein's conviction regarding the incompleteness of quantum mechanics. For if this relation merely reflects the operational constraints of measurement rather than an ontological feature of reality, then quantum theory fails to capture the underlying ontic state. Einstein's stance can thus be characterized as a form of classical universalism, in which quantum superpositions represent nothing more than ``provisional epistemic deficits'' that an authentic objective theory would inevitably eliminate. In this view, superpositions themselves ``conceal'' the ontic state of the physical universe, and the task of a universal classical theory is to uncover the hidden variables that describe this determinate state. However, unless one abandons measurement independence, classical Einsteinian universalism clashes with the constraints of Bell's theorem and must forgo dynamical locality.

The perspective of a quantum-classical bipartite universe, assumed in this work, offers a possible solution to the dynamical non-locality paradox and a fundamentally different account of the role of superpositions. Correlation is maintained not by dynamical non-locality, but because spatially separated, strictly local measurement events are actualizing a joint potentiality within a consistent classical logical framework. There is no ``spooky action at a distance'' because there is no non-local dynamic action, but only logical actualization. Conversely, quantum superpositions possess an ontological character, rendered empirically inaccessible solely by a logical constraint of communicability -- which is equivalent to a physical constraint of commutativity among the physical states (understood as density matrices). It is the Boolean logical constraint that conceals the superpositions, not the superpositions that conceal the ontic state, inverting the Einsteinian premise at the very heart of the measurement problem.

It should be emphasized that this work constitutes a preliminary conceptual framework rather than a complete mathematical formalization. Many technical details are deferred to subsequent investigations. In particular, several avenues for further development can be identified:
\begin{itemize} 
\item the correspondence between the cyclic clock solution and its generator description warrants explicit formulation, particularly in relation to 't Hooft's cogwheel formalism at the Planck scale;
\item the stability analysis requires extension beyond the logical framework to encompass dynamical regimes where gravitational effects become operationally significant;
\item the dual role of $\hbar$ as both a universal constant for quantum dynamics and a scaling parameter for classical clock dynamics demands deeper investigation, especially near Planck-scale energy densities where temporal discreteness becomes operationally relevant;
\item the inversion of the Einsteinian premise -- that Boolean constraints conceal superpositions rather than the superpositions that conceal the ontic state -- demands rigorous engagement with experimental tests of Bell-type correlations and clarification of the ontological status of quantum potentials.
\end{itemize}
Collectively, these directions may advance our logico-physical program toward a comprehensive theory of observation and temporality in a quantum-classical bipartite framework.

\end{document}